\documentclass[10pt,conference]{IEEEtran}
\IEEEoverridecommandlockouts
\usepackage{cite}
\usepackage{amsmath,amssymb,amsfonts}
\usepackage{algorithmic}
\usepackage{graphicx}
\usepackage{textcomp}
\usepackage{xcolor}
\def\BibTeX{{\rm B\kern-.05em{\sc i\kern-.025em b}\kern-.08em
    T\kern-.1667em\lower.7ex\hbox{E}\kern-.125emX}}

\usepackage{tabularx}
\usepackage[table]{xcolor}
\usepackage{booktabs}
\usepackage{todonotes}
\usepackage{hyperref}

\definecolor{lightgray2}{gray}{0.85}
\begin{document}

\title{MARS: Sound Generation via Multi-Channel Autoregression on Spectrograms\\
}

\author{\IEEEauthorblockN{Eleonora Ristori$^{\star \dagger}$}
\IEEEauthorblockA{\textit{AI Lab, DINFO} \\
\textit{Università di Firenze}\\
Florence, Italy \\
eleonora.ristori@unifi.it}
\and
\IEEEauthorblockN{Luca Bindini$^{\star}$\thanks{$^{\star}$These authors contributed equally to this work. $^\dagger$Corresponding author.}}
\IEEEauthorblockA{\textit{AI Lab, DINFO} \\
\textit{Università di Firenze}\\
Florence, Italy \\
luca.bindini@unifi.it}
\and
\IEEEauthorblockN{Paolo Frasconi}
\IEEEauthorblockA{\textit{AI Lab, DINFO} \\
\textit{Università di Firenze}\\
Florence, Italy \\
paolo.frasconi@unifi.it}

% \author{\IEEEauthorblockN{Eleonora Ristori}
}

\maketitle

\begin{abstract}
% Research on audio generation has progressively shifted from waveform-based approaches to spectrogram-based methods, which more naturally capture harmonic and temporal structures. At the same time, advances in image synthesis have shown that autoregression across scales, rather than tokens, improves coherence and detail. However, these two lines of research have largely evolved independently, and their integration remains unexplored. We introduce MARS (Multi-channel AutoRegression on Spectrograms), a framework that treats spectrograms as multi-channel images and employs channel multiplexing (CMX), a reshaping technique that lowers height and width without discarding information. A shared tokenizer provides consistent discrete representations across scales, enabling a transformer-based autoregressor to refine spectrograms from coarse to fine resolutions efficiently. Experiments on a large-scale dataset demonstrate that MARS performs comparably or better than state-of-the-art baselines across multiple evaluation metrics, establishing an efficient and scalable paradigm for high-fidelity sound generation.
Research on audio generation has progressively developed along both waveform-based and spectrogram-based directions, giving rise to diverse strategies for representing and generating audio. At the same time, advances in image synthesis have shown that autoregression across scales, rather than tokens, improves coherence and detail. Building on these ideas, we introduce MARS (Multi-channel AutoRegression on Spectrograms), which, to the best of our knowledge, is the first adaptation of next-scale autoregressive modeling to the spectrogram domain. MARS treats spectrograms as multi-channel images and employs channel multiplexing (CMX), a reshaping strategy that reduces spatial resolution without information loss. A shared tokenizer provides consistent discrete representations across scales, enabling a transformer-based autoregressor to refine spectrograms from coarse to fine resolutions efficiently. Experiments on a large-scale dataset demonstrate that MARS performs comparably or better than state-of-the-art baselines across multiple evaluation metrics, establishing an efficient and scalable paradigm for high-fidelity sound generation.
\end{abstract}


\section{Introduction}

In recent years, audio generation has progressed rapidly, driven by deep generative models that synthesize either raw waveforms or time-frequency representations. 
Despite impressive results, generating long, high-fidelity audio remains challenging due to the need to capture structure at multiple temporal scales while preserving fine detail\cite{DBLP:conf/iclr/KreukSPSDCPTA23, DBLP:conf/icml/LiuCYMLM0P23, DBLP:conf/nips/CopetKGRKSAD23}.

A common distinction is between time-domain and frequency-domain approaches. 
Time-domain models synthesize waveforms directly, including VAE-based methods \cite{peng2020non}, GAN-based architectures \cite{yamamoto2020parallel, DBLP:conf/iclr/BinkowskiDDCECC20}, and diffusion models \cite{DBLP:conf/iclr/KongPHZC21}. 
While effective, these models can face scalability and coherence issues over long horizons, as also highlighted in autoregressive waveform modeling such as MelNet \cite{Vasquez2019MelNetAG}. 
Frequency-domain methods instead generate spectrograms that are later inverted to waveforms, benefiting from explicit harmonic and temporal structure, e.g., GANSynth \cite{DBLP:conf/iclr/EngelACGDR19} and EDMSound \cite{zhu2023edmsound}. 
However, for high-fidelity sound, they must preserve fine spectral detail and support large time-frequency resolutions.

In parallel, image generation has recently seen a shift from raster-scan next-token prediction to next-scale autoregression, where models predict an entire resolution level conditioned on coarser levels. 
Visual AutoRegressive (VAR) modeling \cite{tian2024visual} demonstrated that coarse-to-fine generation can improve global coherence and accelerate inference by reducing sequential depth. 
Subsequent work, such as ImageFolder \cite{DBLP:conf/iclr/0106Q0KGRL25}, further strengthened this paradigm with improved tokenization and training objectives.

Motivated by these developments, we introduce MARS (Multi-channel AutoRegression on Spectrograms), a framework that adapts next-scale autoregression to audio by treating spectrograms as multi-channel images. 
A direct application of VAR-style pipelines to spectrograms is non-trivial: audio spectrograms can be substantially larger than standard image inputs, quickly becoming computationally prohibitive.

To address this issue, we propose channel multiplexing (CMX), a lossless reshaping strategy that reduces the spectrogram spatial size while redistributing information along the channel dimension. 
This makes large time-frequency inputs compatible with architectures designed for smaller spatial grids, enabling efficient training and inference without discarding frequency content.
We further employ a shared tokenizer across resolutions to obtain consistent discrete representations and train a transformer autoregressor to refine spectrograms from coarse to fine.
Our contributions are:
\begin{itemize}
    \item We propose MARS, the first adaptation of next-scale autoregression \cite{tian2024visual} to spectrogram-based sound sample generation.
    \item We introduce CMX, a lossless preprocessing technique that decouples information density from time-frequency resolution by shifting complexity to the channel dimension, thereby substantially reducing the computational cost of both training and inference.
    \item We demonstrate on NSynth \cite{DBLP:conf/icml/EngelRRDNES17} that MARS achieves competitive or superior quality/diversity compared to strong baselines under the evaluation protocol of \cite{DBLP:conf/ismir/Vinay022}.
\end{itemize}
An interactive demo is available at  \href{https://eleonoraristori.github.io/mars-demo/}{eleonoraristori.github.io/\\mars-demo}.

\section{Related Work}
Recent progress in generative modeling suggests two converging trends that motivate our approach. 
First, in image synthesis, discrete tokenization coupled with scale-wise autoregressive generation has improved global coherence while reducing sequential depth. 
Second, in audio and music generation, models increasingly leverage structured representations (waveform, spectrograms, or discrete latents) and evaluate trade-offs between fidelity, diversity, and scalability.
In the following, we summarize the most relevant developments in (i) image generation via discrete representations and next-scale autoregression, and (ii) audio and music generation in both time and frequency domains.

\subsection{Image Generation}
Autoregressive image generation has traditionally been formulated as raster-scan next-token prediction, where each pixel (or token) is generated sequentially.
Representative approaches include PixelRNN \cite{DBLP:conf/icml/OordKK16} and PixelCNN \cite{DBLP:conf/nips/OordKEKVG16}, as well as improved likelihood parameterizations such as PixelCNN++ \cite{DBLP:conf/iclr/SalimansK0K17}. 
While conceptually simple, this approach scales poorly with resolution due to long sequence lengths and limited parallelism.
A key enabler of transformer-based image generation is the use of discrete latent representations, where an encoder maps images to a grid of discrete codes (e.g., vector-quantized tokenizers), allowing transformers to model shorter sequences in token space rather than pixel space.
VQ-VAE \cite{DBLP:conf/nips/OordVK17} is a canonical method in this direction, and later tokenizers such as those used in VQGAN/Taming Transformers \cite{DBLP:conf/cvpr/EsserRO21} further improved perceptual quality at higher resolutions.

More recently, VAR \cite{tian2024visual} proposed a different autoregressive factorization: instead of predicting the next token, the model predicts the next resolution scale. 
Starting from a coarse token map, the model progressively refines the output across a pyramid of resolutions, improving global consistency at low scales while injecting detail at higher scales.
This next-scale strategy reduces the sequential depth of generation to the number of scales, yielding faster inference and stronger coherence.
ImageFolder \cite{DBLP:conf/iclr/0106Q0KGRL25} further improved tokenizer and training design relative to the original VAR pipeline, strengthening the quality of discrete representations and stabilizing learning at high resolutions.
These advances motivate our adaptation of next-scale autoregression to spectrograms, where long time--frequency grids make sequential token prediction particularly expensive.

\subsection{Audio and Music Generation}
Generative modeling for audio and music has been explored in both the time domain and the frequency domain.
Time-domain approaches synthesize raw waveforms directly, using VAEs \cite{peng2020non}, GANs \cite{yamamoto2020parallel, DBLP:conf/iclr/BinkowskiDDCECC20}, and diffusion models \cite{DBLP:conf/iclr/KongPHZC21}. 
Autoregressive waveform models can achieve high fidelity but often face scalability challenges over long horizons \cite{Vasquez2019MelNetAG}.
In contrast, frequency-domain approaches generate spectrograms, leveraging their explicit harmonic structure; representative methods include GANSynth \cite{DBLP:conf/iclr/EngelACGDR19} and diffusion-based spectrogram generators such as EDMSound \cite{zhu2023edmsound}. 
These models shift part of the difficulty to spectrogram inversion and the preservation of fine spectral detail.

A complementary line of work exploits structured audio priors, such as differentiable signal processing, to improve controllability and realism; DDSP \cite{DBLP:conf/iclr/EngelHGR20} is a prominent example. 
Within the NSynth setting \cite{DBLP:conf/icml/EngelRRDNES17}, prior work has also established standardized evaluation protocols and metrics for comparing generation quality and diversity \cite{DBLP:conf/ismir/Vinay022}. 

Our method connects these threads by bringing next-scale autoregression \cite{tian2024visual} to spectrogram generation. 
% Unlike images, spectrograms can be substantially larger for realistic sampling rates and durations; we therefore introduce CMX to reduce spatial resolution losslessly via channel redistribution, and we rely on a shared tokenizer across scales to support coarse-to-fine refinement in a discrete latent space.

\section{Methodology}

Training AR models consists of two stages: first, learning a tokenizer that remains consistent across all resolutions; and second, training the AR model to predict higher-resolution tokens conditioned on the lower-resolution token map.

\begin{figure}[b]
  \centering
   \includegraphics[width=1.0\linewidth]{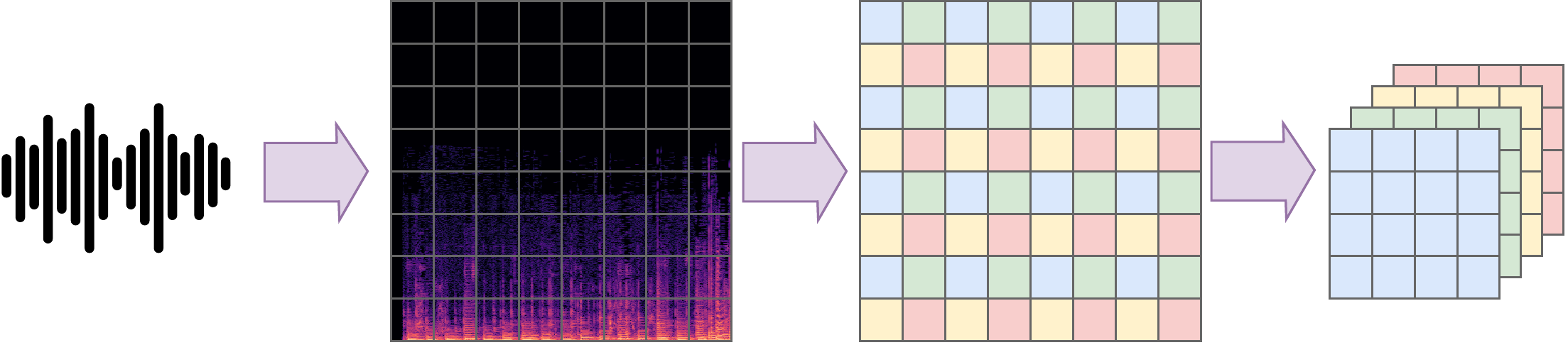}
   \caption{Waveform preprocessing for Tokenizer Input and resolution reduction via Channel Multiplexing (CMX).
   Raw audio files are first transformed into spectrograms using STFT, retaining only the amplitude component. The resulting spectrograms are then reshaped using the CMX technique, which distributes pixel values across multiple channels according to a chessboard-like scheme, as illustrated. This redistribution balances information across channels and is adjusted to match the desired input dimensions (e.g. $256 \times 256 \times C$), where $C$ denotes the number of channels, which depends on the input sampling rate and length.
   }
   \label{fig: preprocessing}
\end{figure}

\begin{figure*}[t]
  \centering
   \includegraphics[width=\linewidth]{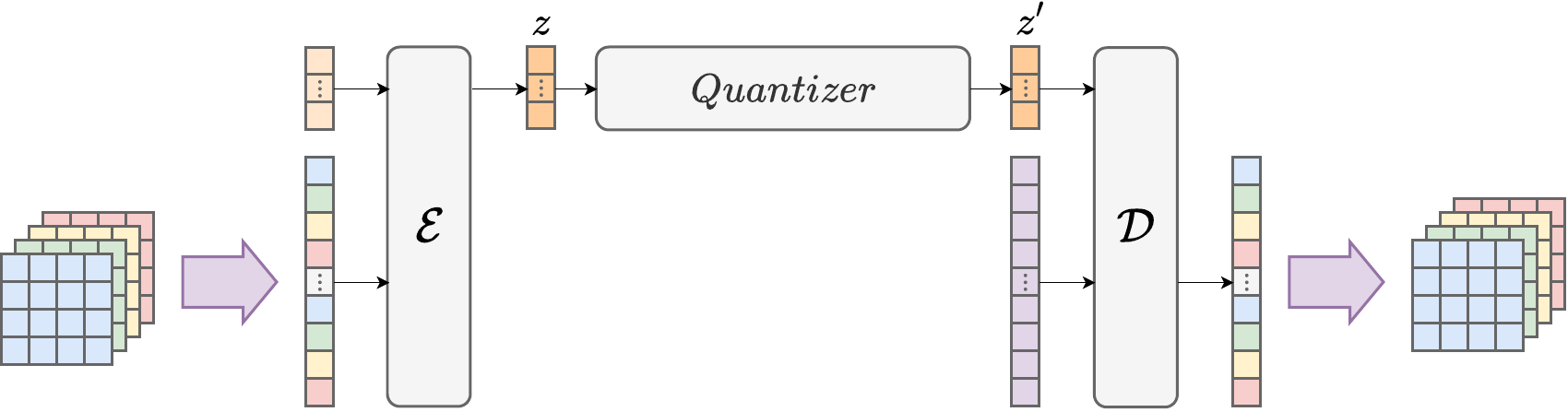}
   \caption{\textit{Tokenizer architecture.} The tokenizer is adapted from ImageFolder \cite{DBLP:conf/iclr/0106Q0KGRL25}, improving upon the original VAR tokenizer \cite{tian2024visual}. The input spectrogram is partitioned into patches of size $L \times L$ and concatenated with $S$ learnable tokens before being processed by a transformer encoder $\mathcal{E}$, producing latent representations $z$. These are discretized by a vector quantizer to obtain $z'$, which are then combined with another set of $L \times L$ learnable tokens and passed to a decoder $\mathcal{D}$ for reconstruction. }
   \label{fig: tokenizer}
\end{figure*}

\subsection{Audio Preprocessing}

We first convert each waveform into a spectrogram using a short-time Fourier transform (STFT), retaining only the amplitude since the phase is later (after inference) reconstructed with the Griffin–Lim algorithm~\cite{griffin}. As an example, an 8s waveform at 16 kHz with a 1024-point STFT and hop size of 256 yields a spectrogram of $512 \times 512$, already matching the largest image size attempted with VAR~\cite{tian2024visual}. Higher sampling rates further enlarge spectrograms, leading to excessive memory consumption. 
Moreover, processing such large inputs inherently increases the number of parameters, and sometimes it even requires greater network depth; as an example, increasing input resolution from $256 \times 256$ to $512 \times 512$ in the original VAR model raised the parameter count from 310M to 2.3B.

To address this problem, which makes our experiments quickly unfeasible, we propose \textit{channel multiplexing} (CMX), which reduces spatial dimensions by redistributing values across channels in a chessboard-like scheme (Figure~\ref{fig: preprocessing}). For spectrograms derived from an $N$-point STFT, the reshaping procedure jointly reorganizes the frequency and temporal dimensions, producing a tensor with the desired reduced spatial resolution and with a channel cardinality $C$ scaling with $N$, the signal length, and sampling rate. In this way, CMX accommodates inputs of varying length and sampling rate by adjusting $C$, while preserving the original data information and substantially reducing spatial overhead. As shown in Section~\ref{effectiveness of CMX}, CMX maintains performance, supports larger batch sizes with lower-variance gradient estimates, and allows the use of architectures designed for smaller inputs—without the need to increase network depth—thereby avoiding significant parameter growth while enabling efficient processing of large-scale data.

% For our tokenizer presented in Figure \ref{fig: tokenizer}, even without increasing network depth, enlarging the input dimension $M \times M$ leads to a parameter growth of $O(K^2+S)$  where $K=M/L$, $L$ is the patch size, and $S$ the number of learnable tokens. This shows that the total number of parameters already scales roughly with the square of the input dimension.
For our tokenizer shown in Figure~\ref{fig: tokenizer}, the number of parameters increases with the input resolution $M \times M$, scaling as $O(K^2 + S)$, where $K = M/L$, $L$ denotes the patch size, and $S$ is the number of learnable tokens. Notably, this growth occurs even without increasing the network depth, indicating that the total parameter count already scales approximately quadratically with the input dimension.

Crucially, the number of parameters is almost independent of the number of input channels. The only layers affected are the initial projection that maps the input into the latent space and the last one that projects it back, which is the only one used at inference time, since the encoder $\mathcal{E}$ is discarded. In fact, there is no need to increase the latent space dimension when processing larger inputs because the spectrogram domain is inherently highly structured, and even for longer signals or with higher frequency resolution, the patterns remain highly repetitive: harmonics, recurring temporal structures, and correlated frequency bands create strong redundancy across both time and frequency axes. This means that the latent space can efficiently capture the essential spectral and temporal information for longer and higher-quality audio without increasing its dimension. 
% This redundancy allows the latent space to efficiently capture the essential spectral and temporal information without increasing its dimensionality. 
Consequently, reducing the input dimension 
$M$ provides substantial benefits, as parameter count scales quadratically with $M$, while still preserving all critical information for high-fidelity audio generation.

More generally, CMX decouples information density from time–frequency resolution. Rather than discarding or compressing data, it reorganizes it into a format naturally suited to architectures optimized for multi-channel inputs (e.g., RGB images). This enables long recordings or high-fidelity signals to be processed with bounded memory cost, yielding a compact, lossless representation that scales efficiently, reducing memory and compute requirements while maintaining full data fidelity.

\subsection{Tokenizer}

The tokenizer architecture and training procedure are adapted from ImageFolder \cite{DBLP:conf/iclr/0106Q0KGRL25}, extending the original VAR tokenizer \cite{tian2024visual}. The overall design is illustrated in Figure \ref{fig: tokenizer}.  

The input multiplexed spectrogram is first partitioned into $L \times L$ patches. These patches are concatenated with a set of $S$ learnable tokens and processed by a transformer encoder $\mathcal{E}$, which produces latent representations $z$. A vector quantizer then discretizes these embeddings into codebook entries $z'$. The quantized tokens are further concatenated with another set of $L \times L$ learnable tokens and passed through a decoder $\mathcal{D}$ to reconstruct the spectrogram.  

The tokenizer is trained using a composite objective that combines three terms:
\begin{equation}
    \mathcal{L} = \lambda_{recon}\mathcal{L}_{recon} + \lambda_{VQ}\mathcal{L}_{VQ} + \lambda_{ad}\mathcal{L}_{ad}.
\end{equation}
Here, $\mathcal{L}_{recon}$ is an $L_2$ reconstruction loss that ensures fidelity between the reconstructed and ground-truth spectrograms; $\mathcal{L}_{VQ}$ is the vector quantization loss, which aligns the encoder outputs with the nearest codebook entries; and $\mathcal{L}_{ad}$ is an adversarial loss applied through a PatchGAN discriminator \cite{isola2017image}, encouraging reconstructions to be indistinguishable from real spectrograms.

\subsection{Autoregressive Model}
Once the tokenizer is trained, we adopt the next-scale autoregressive paradigm introduced by Visual AutoRegressive (VAR) modeling~\cite{tian2024visual}. Unlike raster-scan next-token prediction, VAR generates the output in a coarse-to-fine manner, predicting entire resolution levels progressively.

Concretely, for each spectrogram we consider a pyramid of discrete token maps
$\{r_1, r_2, \dots, r_K\}$, where each $r_k$ is a token grid at a given resolution and all tokens belong to the same vocabulary (shared tokenizer/codebook across scales). The model is trained to predict each finer-scale representation conditioned on coarser ones, i.e., it learns a distribution of the form
\begin{equation}
p(r_1,\dots,r_K) = \prod_{k=1}^{K} p(r_k \mid r_{<k}),
\label{eq:var_factorization}
\end{equation}
where $r_{<k}$ denotes the set of previously generated (coarser) token maps.

We implement this idea with a decoder-only transformer. Following Tian et al.~\cite{tian2024visual}, training is performed by presenting the model with the coarser scales and training it to predict the next scale. A block-wise causal attention mask enforces the intended dependency structure: tokens at a given scale can attend to the conditioning token and to all tokens from earlier (coarser) scales, but not to future (finer) scales. In this way, the model captures global structure at low resolutions and refines details as resolution increases, while keeping the computation manageable.
This strategy is especially effective for spectrograms, whose time–frequency representations are often high-dimensional, commonly reaching $4\times$ to $8\times$ the size of conventional $256\times256$
image inputs: a raster-scan autoregressive transformer would require a very large number of sequential steps, whereas next-scale prediction reduces the sequential depth of generation to the number of scales. In our setting, CMX further helps by bounding the spatial resolution while preserving information in the channel dimension, making the VAR-style pipeline practical for audio.

\subsection{Inference}
At inference time, we generate sound by sampling the multi-scale token pyramid from coarse to fine, then decoding tokens back to a magnitude spectrogram and reconstructing the waveform.

Generation starts from a conditioning token $[s]$ and proceeds across scales $k=1,\dots,K$. At each stage, the transformer takes as context the already generated coarser token maps and produces a prediction for the next scale. Importantly, tokens within the same scale can be sampled in parallel, so each refinement stage typically requires only a small number of forward passes (and can benefit from standard transformer optimizations such as KV-caching~\cite{tian2024visual}).

After sampling the finest token map $r_K$, we decode it through the tokenizer decoder $\mathcal{D}$ to obtain a multiplexed magnitude spectrogram. We then apply the inverse CMX reshaping to recover the original time--frequency layout. The waveform is finally obtained by applying the Griffin-Lim algorithm\cite{griffin}.

\begin{table*}
  \caption{Results according to the objective metrics reported in~\cite{DBLP:conf/ismir/Vinay022}. The best results are highlighted in bold, and the second best are underlined. The last (highlighted) column reports metrics on generated sound (i.e., after the autoregressive module), and values are not directly comparable against those in the previous four columns.}
\label{table:res_1}
\small
\centering
\begin{tabularx}{0.7\textwidth}{l|*{4}{>{\centering\arraybackslash}X|}|>{\centering\arraybackslash\columncolor{lightgray2}}X}
\toprule
Metric & Diffwave & DDSP & NSynth & \textbf{MARS (ours)} & \textbf{MARS (generated)} \\
\midrule
NDB/$k$ ($\downarrow$) & 0.74 & \underline{0.20} & 0.74 & \textbf{0.19} & 0.15 \\
PKID ($\downarrow$) & 0.0093 & \underline{0.0053} & 0.0101 & \textbf{0.0035} & 0.0066 \\
IKID ($\downarrow$) & 0.0021 & \underline{0.0020} & 0.0024 & \textbf{0.0015} & 0.0017 \\
PIS ($\uparrow$) & 2.3814 & \textbf{3.3224} & 2.3238 & \underline{2.9602} & 3.2220 \\
IIS ($\uparrow$) & \textbf{5.6477} & \underline{5.3371} & 4.6364 & 5.2047 & 4.5357 \\
MSE ($\downarrow$) & 0.0291 & \textbf{0.0130} & 0.0329 & \underline{0.0143} & 0.0201 \\
MAE ($\downarrow$) & 0.1369 & \textbf{0.0666} & 0.1224 & \underline{0.0915} & 0.0795 \\
FAD ($\downarrow$) & 7.9488 & \textbf{1.1519} & 4.0590 & \underline{1.6429} & 1.8833 \\
\bottomrule
\end{tabularx}
\end{table*}

\section{Experiments}
In this section, we present an experimental evaluation of our approach on the NSynth dataset, analyzing both reconstruction and generation performance. The goal is to assess audio quality, sample diversity, and perceptual fidelity under the standardized evaluation protocol proposed by  \cite{DBLP:conf/ismir/Vinay022}.

\subsection{Experimental Setup} We evaluate our autoregressive model on the NSynth dataset \cite{DBLP:conf/icml/EngelRRDNES17}, which contains over 300,000 monophonic musical notes characterized by distinct pitch, timbre, and envelope properties. Each audio sample has a duration of approximately 4 seconds and is recorded at a sampling rate of 16 kHz. We convert the waveforms into spectrograms using a 1024-point short-time Fourier transform (STFT) and rearrange them with CMX into tensors of size $256 \times 256 \times 2$, where frequency components are split across two channels.

% The image encoder, following the DINOv2-base architecture, is initialized with random weights. 
The tokenizer’s transformer encoder follows a DINOv2-base architecture and it's initialized with random weights. We adopt a cosine learning rate scheduler with a warmup for 1 epoch
and a start learning rate of 3e-5.
The tokenizer employs a codebook of size 16,384. We train the tokenizer for 400 epochs following the training protocol of \cite{DBLP:conf/iclr/0106Q0KGRL25}, with loss weights set to $\lambda_{recon}=\lambda_{VQ}=1$ and $\lambda_{ad}=0.5$. We set the tokenizer residual scales to [1, 1, 2, 3, 3, 4, 5, 6, 8, 11]. The autoregressive model is then trained for 350 epochs using the settings described in \cite{tian2024visual}. All experiments are conducted on a single RTX 5000 Ada GPU with 32 GB of memory.

\subsection{Evaluation Metrics}
To assess our model, we follow the comprehensive evaluation protocol proposed by \cite{DBLP:conf/ismir/Vinay022}, which benchmarks state-of-the-art audio generation methods on the NSynth dataset. In line with this methodology, we evaluate our tokenizer for audio reconstruction using a diverse set of metrics that capture complementary aspects of audio. These metrics are grouped into three categories:

\begin{itemize}
    \item \textit{Reconstruction errors}. We report Mean-Squared-Error (MSE) and the Mean-Absolute-Error (MAE), computed on mel-spectrograms extracted from audio samples, as recommended in \cite{DBLP:conf/ismir/Vinay022}.
    \item \textit{Sample diversity metrics}. 
    The number of statistically different bins (NDB/$k$) \cite{richardson2018gans} quantifies sample diversity. It is computed via a Voronoi decomposition of $k$-means clusters in the training space. Test samples are assigned to clusters using $L_2$ distance, and a two-sample $t$-test is used to identify statistically different bins. The final score is the fraction of bins that differ significantly. An NDB/$k$ score close to 0 indicates good diversity and coverage of the training distribution, whereas values near 1 suggest mode collapse. 
    % The Number of Statistically Different Bins (NDB/$k$) \cite{richardson2018gans} measures sample diversity by clustering the training space and comparing the distribution of test samples to training data. Scores near 0 indicate good coverage of the training distribution, while values close to 1 reveal mode collapse.
    %
    To further assess sample diversity and semantic consistency, we adopt the Pitch Inception Score (PIS) and Instrument Inception Score (IIS) \cite{nistal2021comparing}, which measure how well reconstructed samples match the true distributions of pitch and instrument classes in the NSynth dataset. These metrics are computed analogously to the standard Inception Score \cite{DBLP:conf/nips/SalimansGZCRCC16}, but rely on two task-specific backbone networks: one trained for pitch classification and the other for instrument classification.
    Following~\cite{DBLP:conf/ismir/Vinay022}, we also report these metrics, although they focus on ``symbolic'' properties of notes rather than audio quality.
    
    \item \textit{Embedding-based metrics}. Kernel Inception Distance (KID) \cite{DBLP:conf/iclr/BinkowskiSAG18} measures the similarity between the distributions of reference and reconstructed embeddings. Specifically, we compute PKID and IKID \cite{nistal2021comparing} using the same Inception networks employed for PIS and IIS, where the maximum mean discrepancy (MMD) is applied to compare the embedding distributions. 
    
    We also adopt Fréchet Audio Distance (FAD) \cite{DBLP:conf/interspeech/KilgourZRS19}, a widely adopted metric for perceptual quality, which leverages VGGish embeddings of reference and reconstructed audio. Each embedding set is modeled as a multivariate Gaussian distribution, $\mathcal{N}_o$ for the original and $\mathcal{N}_r$ for the reconstructed samples, and FAD is defined as the Fréchet distance between these two distributions.
\end{itemize}
 In addition, we also evaluate our AR model for generating audio using the same metrics. MSE and MAE in this case are calculated considering for each generated sample the closest spectrogram in the NSynth test set.

\subsection{Experimental Results}
Table~\ref{table:res_1} compares our MARS with NSynth \cite{DBLP:conf/icml/EngelRRDNES17}, DDSP \cite{DBLP:conf/iclr/EngelHGR20}, and DiffWave \cite{DBLP:conf/iclr/KongPHZC21}. MARS achieves the best scores in NDB/$k$, PKID, and IKID, indicating superior sample diversity and fidelity in both pitch and timbre. It also ranks among the top methods for MSE, MAE, and FAD, reflecting accurate reconstruction and high perceptual quality. PIS and IIS, which measure similarity on sound properties rather than audio quality, remain competitive. 

The last column reports metrics on generated audio, which are not directly comparable to the reconstruction metrics; nonetheless, the generated samples maintain low reconstruction error and perceptual similarity, confirming that MARS effectively preserves signal characteristics during synthesis.

\subsection{Effectiveness of CMX in Maintaining Audio Quality}
\label{effectiveness of CMX}
In this section, we present a toy example evaluating the impact of CMX. We trained the tokenizer of our AR model for 200k steps using two setups: (i) truncated spectrograms of size $256\times256\times1$ obtained from original $512 \times 512 \times 1$, and (ii) truncated spectrograms rearranged with CMX into  $128\times128\times4$. Training with CMX  reduced training time by $1.5\times$, while still achieving reconstruction quality comparable to the truncated baseline. CMX outperformed truncation in terms of reconstruction accuracy, with lower errors (MSE improved by 0.002 and MAE by 0.022), but showed a slight drop in perceptual quality, as indicated by a 0.19 increase in FAD. Crucially, CMX enables the preservation of all frequency information, which is essential for audio quality, while making this feasible through its substantial memory reduction. Indeed, frequency truncation leads to severe degradation (FAD gap of 2.16 and MSE gap of 0.01 compared to our experiments on input of dimension $256 \times 256 \times 2$), confirming that CMX is necessary to balance efficiency with fidelity. Overall, these results demonstrate that CMX reduces memory costs, improves reconstruction fidelity, and preserves frequency resolution, with only a marginal trade-off in perceptual quality.

% \subsection{Future Work}
% % While our current experiments are limited to the NSynth dataset, which serves as a standard benchmark for generative audio and enables comparison under the evaluation protocol of \cite{DBLP:conf/ismir/Vinay022}, we consider this study only a first step. 

% Unlike many recent approaches that focus on end-to-end generation of full songs—often at limited perceptual quality—conditioned on textual prompts or visual inputs, our objective is fundamentally different. We aim to generate short, high-fidelity audio samples that can be directly used by musicians and audio engineers as building blocks in professional music production workflows.

% In future work, we plan to extend our evaluation to more complex and diverse music datasets featuring richer instrumentation, longer temporal contexts, and more intricate harmonic structures, as well as higher sampling rates, such as 48 kHz high-fidelity audio. We also intend to explore applications in the speech domain, where accurately modeling fine temporal details and long-range dependencies poses additional challenges. Expanding to these domains will allow us to further assess the generality and scalability of MARS and CMX, and to investigate their potential for high-quality audio generation in real-world creative and technical settings.

\section{Conclusion}
In this work, we introduced MARS (Multi-channel AutoRegression on Spectrograms), a novel framework for sound sample generation that adapts next-scale autoregression from image synthesis to the spectrogram domain. By introducing the channel multiplexing (CMX) technique, MARS reduces spatial resolution while preserving frequency information, enabling scalable training with a manageable memory footprint. The same principle
can be extended beyond audio to images, video, medical scans, or structured time series since CMX  provides a compact,
lossless representation that scales efficiently, reducing memory
and compute requirements while maintaining full data fidelity. Within the MARS framework, using a shared tokenizer across scales further ensures consistent discrete representations, allowing the autoregressive model to refine spectrograms hierarchically. 

The experimental evaluation on the NSynth dataset demonstrated that MARS achieves competitive or superior performance compared to state-of-the-art baselines across multiple metrics, confirming both the effectiveness of CMX in balancing efficiency with fidelity and the robustness of the overall framework for high-quality audio generation.

Unlike approaches targeting end-to-end generation of complete songs (often at limited perceptual quality) from high-level conditioning, our objective is to generate  high-fidelity audio samples that can be directly used as building blocks in professional music production workflows. As future work, we plan to extend our evaluation to more complex and diverse music datasets with richer instrumentation, longer temporal contexts, and higher sampling rates (e.g., 48 kHz). We also intend to investigate applications in the speech domain, where modeling fine temporal detail and long-range dependencies is particularly challenging, to further assess the generality and scalability of MARS and CMX in real-world settings.

\bibliographystyle{IEEEtran}
\bibliography{refs}

\end{document}